\documentclass[prd,a4paper,showpacs,byrevtex,preprint]{revtex4}
\usepackage{graphicx}
\usepackage{dcolumn}
\usepackage{amsmath}
\usepackage{array}
\usepackage{bm}
\usepackage{amssymb}
\usepackage{amsfonts}

\begin{document}
\title{Effective potential between two transverse gluons from lattice QCD}

\author{Fabien \surname{Buisseret}}
\email[E-mail: ]{fabien.buisseret@umh.ac.be}
\affiliation{Groupe de Physique Nucl\'{e}aire Th\'{e}orique,
Universit\'{e} de Mons-Hainaut,
Acad\'{e}mie universitaire Wallonie-Bruxelles,
Place du Parc 20, BE-7000 Mons, Belgium}

\date{\today}

\begin{abstract}
Modeling glueballs as bound states of transverse constituent gluons allows to understand the main features of the lattice QCD glueball spectrum. In particular it has been shown in previous works that the lightest $C$-even glueballs can be seen as bound states of two transverse constituent gluons interacting \textit{via} a funnel potential. In the present study we show that such an effective potential emerges from the available lattice QCD data. Starting from the scalar glueball mass and wave function computed in lattice QCD, we indeed compute the equivalent local potential between two transverse constituent gluons in the scalar channel and show that it is compatible with a funnel shape, where standard values of the parameters are used and where a negative constant has to be added to reproduce the absolute height of the potential. Such a constant could be related to instanton-induced effects in glueballs. 
\end{abstract}



\maketitle

\section{Introduction}
The study of glueballs is an active domain of hadronic physics, both experimentally and theoretically. Much experimental effort has been devoted to the identification of a clear glueball signal, but no unambiguous candidate has been found up to now~\cite{review}. It is nevertheless generally accepted that a state with a dominant glueball component should be present among the many known $f_0$ resonances~\cite{Ochs}. From a theoretical point of view, a remarkable achievement has been the computation of the glueball spectrum thanks to lattice QCD calculations~\cite{lattice1,latti4}. Many works have since been devoted to the understanding of this spectrum within different frameworks: Coulomb gauge QCD~\cite{szcz}, AdS/QCD duality~\cite{ads}, potential models~\cite{glum0,glumm}, \textit{etc} -- see also Ref.~\cite{vrev} for a review.

In potential models, there are two basic ingredients. First, one assumes that glueballs are bound states of constituent gluons. Second, one introduces an interaction potential that aims to describe at best the QCD interactions. Let us discuss these two points. 

As it is shown in Ref.~\cite{glub}, the structure of the whole lattice QCD glueball spectrum can be understood by assuming that a particular $J^{PC}$ glueball can be labeled by a given number of \textit{transverse} (helicity-1) constituent gluons. The low-lying $C=+$ ($-$) states are then seen as mainly two- (three-) gluon states. The problem of using longitudinal (spin-1) constituent gluons is that light $1^{\pm+}$states are then present in the spectrum, in disagreement with lattice QCD~\cite{glum0,glumm}. However, if transverse constituent gluons are used, the spurious states disappear~\cite{szcz}, and a simple potential model (two-body spinless Salpeter Hamiltonian with funnel potential $ar-\kappa/r$) becomes able to reproduce with accuracy the results of lattice QCD in the $C=+$ sector~\cite{gluheli}. Even if the transversality is assumed, the value of the gluon mass is still a matter of discussion. There are works in the spirit of Refs.~\cite{glum0} arguing that a constituent gluon is massless at the level of the confining Hamiltonian (zero bare mass) but then gains a constituent mass because of its confinement into a glueball. In this picture, the constituent gluon is \textit{a posteriori} massive and its constituent mass is typically given by $\left\langle \sqrt{\bm p^2}\right\rangle$ -- notice that the same argument would apply for $u$ and $d$ quarks. In Ref.~\cite{lhade} for example, the constituent gluon mass is computed to be around 300~MeV. Other studies~\cite{glumm} rather state that a constituent gluon is \textit{a priori} massive, that is with a nonzero bare mass. The underlying idea is roughly that the nonperturbative effects of QCD cause a mass term to appear in the gluon propagator. In the present work, motivated by the results of Refs.~\cite{gluheli,glub}, we will only focus on the case where the constituent gluons have a vanishing bare mass. 

The potential term remains to be chosen. The potential energy between a heavy quark-antiquark pair has been accurately computed in lattice QCD and exhibits a clear funnel shape at the dominant order~\cite{lat0}, supplemented with the usual Fermi-Breit relativistic corrections~\cite{koma}. The potential energy between two static sources in color octet (static gluons) has also been computed in lattice QCD; it is still compatible with a funnel form, for different values of the parameters because static gluons are color-octet sources rather than triplet ones~\cite{lat0}. But, constituent gluons should be seen as relativistic particles and one can wonder if such a static potential is relevant to describe glueballs. We have recently proposed in Ref.~\cite{glue2} a new method to extract the effective potential between two ``physical" gluons (in opposition with static ones) from lattice QCD. The starting point is that not only glueball masses can be computed on the lattice, but also glueball wave functions, especially in the scalar channel~\cite{lat,gluwf}. Assuming that the scalar glueball is a bound state of two constituent gluons, the inverse problem can be solved, \textit{i.e.} to compute the effective potential corresponding to the lattice QCD glueball mass and wave function from a spinless Salpeter-type Hamiltonian. We have given a detailed resolution of this problem in Ref.~\cite{glue2}, but longitudinal constituent gluons were used. Motivated by the results of Refs.~\cite{szcz,glub,gluheli} favoring transverse constituent gluons, we propose in the present work to reconsider the calculations of Ref.~\cite{glue2} and see whether the effective potential that emerges is compatible with a funnel one or not in the case of transverse constituent gluons. 

Our paper is organized as follows. We first recall the lattice QCD data that we have at our disposal in Sec.~\ref{latti}. Then we discuss the expected form of a Hamiltonian describing a bound state of two transverse gluons in Sec.~\ref{hami} and we focus on the scalar channel, for which reliable lattice data are currently available. In Sec.~\ref{numres} we numerically compute the effective gluon-gluon potential and comment its compatibility with a standard funnel form and with the set of parameters used in Ref.~\cite{gluheli}. Finally, we draw some conclusions in Sec.~\ref{conclu}.

\section{Results from lattice QCD}\label{latti}

General field-theoretical arguments have been invoked to prove that the lightest glueball is a scalar one, \textit{i.e.} with quantum numbers $0^{++}$~\cite{0pp}. This feature has been confirmed by several lattice QCD studies, where the most recent determination of the $0^{++}$ glueball mass is around $1.7$~GeV~\cite{lattice1,tepernew}.  

Glueball masses are not the only observables that can be obtained within the framework of lattice QCD: Wave functions can also be computed. A first determination of glueball radial wave functions in the scalar and tensor channels has been made in Ref.~\cite{lat} within SU(2) lattice QCD simulations, while an SU(3) generalization of these results has been given in Ref.~\cite{gluwf}. In this last reference, it is found that
\begin{equation}\label{p1}
m_{0^{++}}=1.680\pm0.046\ {\rm GeV},
\end{equation}
in agreement with Ref.~\cite{lattice1}. It can be checked in Fig.~\ref{fig2} that the lattice radial glueball wave function in the scalar channel is well fitted by the following expression~\cite{glue2}
\begin{equation}\label{fit}
  R(r)=2.5\, \exp\left[-A\left(r/r_0\right)^B\right],
\end{equation}
\begin{eqnarray}\label{p2}
 {\rm with}\quad  r_0&=&1.472\ {\rm GeV}^{-1},\nonumber \\
  A&=&0.883\pm0.045,\quad  B=1.028\pm0.132.\nonumber
\end{eqnarray}
These results have been confirmed in the more recent work~\cite{loannew}. Notice that $R(r)$ has been here normalized in such a way that $R(0)=2.5$ so that the plot is clearer in Fig.~\ref{fig2}, but the particular normalization of the wave function is irrelevant for the computations that we make in the following.  

As it is argued in Ref.~\cite{glue2}, the lattice wave function~(\ref{fit}) can be seen as the Bethe-Salpeter wave function of a stationary scalar state made of two constituent gluons, in the rest frame of the system. It is thus relevant to identify the lattice wave functions of Refs.~\cite{lat,gluwf} with the Schr\"{o}dinger-like wave functions of two-gluon bound states. We refer the reader to Ref.~\cite{glue2} for a detailed discussion of that point.  

\section{Hamiltonian for two transverse gluons}\label{hami}

The problem we want to solve in this work is to find a Hamiltonian $H$ such that
\begin{equation}\label{eigen}
  H\, R(r)=m_{0^{++}}\, R(r),
\end{equation}
where $R(r)$ and $m_{0^{++}}$ come from lattice QCD, as recalled in the previous section. Since there are strong evidences supporting the identification of the lowest-lying scalar gluleball as a mainly two-gluon bound state~\cite{glum0,glumm,glub,gluheli}, we choose for $H$ a standard two-body spinless Salpeter form, i.e. 
\begin{equation}\label{hamdef}
H=2\ \sqrt{\bm p^{\, 2}}+V(r).
\end{equation}
We assume the constituent gluons to be transverse and massless as in Ref.~\cite{gluheli} and the potential $V(r)$ to be local and radial. Note also that $\bm p^{\, 2}=p^2_r+ \bm L^{\, 2}/r^2$.

Once the radial quantum number $n$ and the matrix element of the square orbital angular momentum $\left\langle \bm L^{\, 2}\right\rangle$ are specified ($n=0$ here since we deal with the ground state), the only unknown quantity in Eq.~(\ref{eigen}) is the local central potential $V(r)$. As we will show in the next section, such a particular inverse problem can be numerically solved with accuracy. But, before making explicit calculations, one has to wonder about the value of $\left\langle \bm L^{\, 2}\right\rangle$. In our previous work~\cite{glue2}, we considered, as it is the case in most of the potential models, that the constituent gluons were longitudinal. In this case, the lowest-lying scalar glueball is a $\left|^1S_0\right\rangle$ state in spectroscopic notation, thus with $\left\langle \bm L^{\, 2}\right\rangle=0$ and $\left\langle \bm S^{\, 2}\right\rangle=0$. However, this assumption neglects the fact that constituent gluons should actually be transverse. We showed in Ref.~\cite{gluheli} that a proper inclusion of the gluon's helicity can be performed within the helicity formalism, first presented in Ref.~\cite{jaco}. The results are then considerably improved: The spectrum that is found is in better agreement with lattice QCD, and spurious states due to longitudinal degrees of freedom disappear. Following the helicity formalism, the unique scalar glueball state corresponds to the quantum state~\cite{gluheli}
\begin{equation}\label{scala}
	\left|0^{++}\right\rangle=\sqrt{\frac{2}{3}}\ \left|^1S_0\right\rangle+\sqrt{\frac{1}{3}}\ \left|^5D_0\right\rangle,
\end{equation}
for which it is readily found that $\left\langle \bm L^{\, 2}\right\rangle=\left\langle \bm S^{\, 2}\right\rangle=2$. Notice that the unique pseudoscalar state is simply $\left|0^{-+}\right\rangle=-\left|^3P_0\right\rangle$~\cite{gluheli}, with also $\left\langle \bm L^{\, 2}\right\rangle=2$. It is worth pointing out that the particular form of~(\ref{scala}) is model-independent: Only the requirement that the total spin $\bm J^{2}$ and $J_z$ are good quantum numbers and that the gluons are transverse is needed to build the spin-angular part of the various two-gluon helicity states from the general formalism of Ref.~\cite{jaco}. 

In summary, once transverse constituent gluons are assumed, the problem that remains to be solved is the inverse problem corresponding to Hamiltonian~(\ref{hamdef}) in the scalar channel, that is
\begin{equation}\label{hamdef2}
H=2\ \left.\sqrt{\bm p^{\, 2}}\right|_{\left\langle \bm L^{\, 2}\right\rangle=2}+V(r).
\end{equation}
One could have thought that, since explicitly spin-dependent terms are only present at the level of the relativistic corrections, the transversality of the constituent gluons would not strongly affect the features of the system. However, the difference between spin-1 and helicity-1 gluons dramatically affects the Hamiltonian of the system even at the dominant order, because $\left\langle \bm L^{\, 2}\right\rangle$ is equal to 2 instead of 0. 

\section{Numerical results}\label{numres}

Several numerical methods exist to compute the equivalent local potential from a
given wave function and its corresponding energy but, to our knowledge, only the Lagrange mesh method is able to make accurately such computations with a semirelativistic
kinematics of the form $\sqrt{\bm p^{\, 2}}$~\cite{lag1}. We refer the reader to Refs.~\cite{lag1,baye} for further elements about this numerical technique, but nevertheless recall the main points for self-completeness.

In the case of radial equations, a Lagrange mesh is formed of $N$ mesh points
$x_i$ which are the zeros of the Laguerre polynomial $L_N(x)$ of degree $N$ (here $N=100$). The Lagrange basis is then given by a
set of $N$ regularized orthonormalized Lagrange functions satisfying the condition $f_j(x_i) \propto \delta_{ij}$. These functions can be expressed in terms of $L_N(x)$. The matrix elements $T_{ij}$ of the operator $\sqrt{\bm p^{\, 2}}$, depending on $\left\langle \bm L^{\, 2}\right\rangle$, can be
accurately computed in this basis. Moreover, for a local radial potential, one has $V_{ij}=V(hx_i)\delta_{ij}$ in the Lagrange basis, with $h$ an appropriate scale parameter. It is readily checked that this nice feature allows to rewrite Eq.~(\ref{eigen}) as $V(hx_i)=f\left[T_{ij},m_{0^{++}},R(hx_i)\right]$, thus to solve the inverse problem by computing the potential at the mesh points~\cite{lag1}. In our case, both $m_{0^{++}}$ and $R(hx_i)$ are known thanks to the lattice results~(\ref{p1}) and (\ref{fit}). 

\begin{center}
\begin{figure}[t]
  \resizebox{0.5\textwidth}{!}{%
  \includegraphics{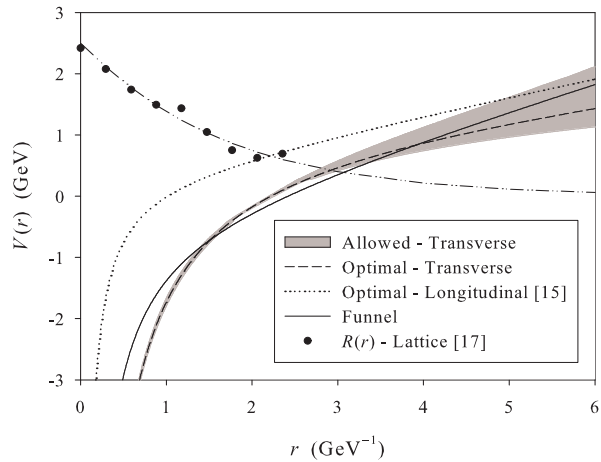}
  }
\caption{Plot of the numerically computed optimal potential $V(r)$ from Eq.~(\ref{hamdef2}) (dashed line). It is computed from the optimal wave function~(\ref{fit}) (dashed-dotted line) fitted on te lattice data of Ref.~\cite{gluwf} (circles). The errors on the
glueball mass $m_{0^{++}}$ and on the wave function parameters $A$ and
$B$ actually allow every potential which is located in the gray area.
These results are compared with the funnel potential~(\ref{corn}) for the values~(\ref{params}) of the parameters (solid line). The optimal potential obtained in Ref.~\cite{glue2} with longitudinal gluons is also plotted for comparison (dotted line). }
\label{fig2}
\end{figure}
\end{center}

Formally, the numerical determination of $V(r)$ in Eq.~(\ref{eigen}) is identical to the one performed in Ref.~\cite{glue2}. The crucial difference however is that $\left\langle \bm L^{\, 2}\right\rangle=2$ this time and not zero as in Ref.~\cite{glue2}. The potential obtained by using the optimal values of $m_{0^{++}}$, $A$, and
$B$ is plotted in Fig.~\ref{fig2}. It exhibits a confining long-range part, and a quickly
decreasing short-range part. The errors on these three parameters [see
Eqs.~(\ref{p1}) and (\ref{p2})] allow the ``true" potential to be
located between two extremal curves, within the shaded area of Fig.~\ref{fig2}. It can clearly be observed that the optimal potential obtained in Ref.~\cite{glue2}, that is with longitudinal gluons, is different from the new one even if the error bars are included. Again, it should be stressed that the use of transverse gluons considerably affects the physical features of the two-gluon system. 

These numerical results are qualitatively compatible with a funnel potential of the form
\begin{equation}\label{corn}
  V_f(r)={\cal C}\, \sigma\,r-\frac{3\alpha_S}{r}-D.
\end{equation}
In this expression, $\alpha_S$ is the strong coupling constant, $\sigma$ is the fundamental quark-antiquark flux tube energy
density, and ${\cal C}$ indicates the scaling of the energy density,
which is different for a gluon-gluon system or a quark-antiquark pair. Several approaches support the Casimir scaling hypothesis, \textit{i.e.} ${\cal C}=9/4$~\cite{cas}. The constant $D$ can be used to fit the height of potential~(\ref{corn}) on the numerically computed optimal potential. 

Instead of fitting the parameters of potential~(\ref{corn}) on our numerical results, we find more interesting to check if they are in agreement with the potential of Ref.~\cite{gluheli}, that leads to a nice agreement with lattice QCD in the $C=+$ sector. This last potential is such that 
\begin{eqnarray}\label{params}
{\cal C}&=9/4,\quad \sigma&=0.185~{\rm GeV}^2,\\
\alpha_S&=0.45,\quad D&=0.450~{\rm GeV}\nonumber.	
\end{eqnarray}
We argued in Ref.~\cite{gluheli} that this particular value of $D$ actually comes from instanton-induced forces in glueballs, that roughly contribute to the mass as $-D\, P\, \delta_{J,0}$~\cite{gluheli}. A look at Fig.~\ref{fig2} shows that the funnel potential~(\ref{corn}) with the parameters~(\ref{params}) is compatible with the numerically computed potential. We notice that the $0^{++}$ mass we find by solving Hamiltonian~(\ref{hamdef}) with $V(r)=V_f(r)$ and with the parameters~(\ref{params}) is $1.724$~GeV, in agreement with the glueball mass~(\ref{p1}).       

\section{Conclusions}\label{conclu}
In this work, we have computed the effective potential between two transverse, massless, constituent gluons from the mass and wave function of the $0^{++}$ glueball obtained in lattice QCD. We followed the method described in Ref.~\cite{glue2}, but this time the helicity of the constituent gluons has been taken into account. The numerically computed potential is compatible with a funnel form with the values of the parameters that we used in Ref.~\cite{gluheli} to reproduce at best all the low-lying $C=+$ glueballs (not only the scalar one). The present results thus draw a bridge between lattice QCD and potential models, and show how physical informations about the relevant effective potential to use can be derived from lattice QCD, especially in the light hadron sector.  

A negative constant is needed to lower the funnel potential to the value that is numerically computed. Such a negative constant has been argued to come from instanton-induced interactions in glueballs in Ref.~\cite{gluheli}. It would be very interesting to have at our disposal the glueball wave function in the $0^{-+}$ channel since in this case the instanton-induced mass term is expected to be positive. Thus we should have $D\approx-0.45$~GeV in the pseudoscalar case, the other parameters being unchanged. Moreover, in the other channels (the tensor one in particular), one would expect $D\approx 0$. New accurate lattice QCD simulations could be very useful to check this point. 

Finally, let us make some comments about the tensor glueball, for which a wave function has also been computed in Refs.~\cite{lat,gluwf}. If constituent gluons were spin-1 particles, the lowest-lying $2^{++}$ state would be a $\left|^5 S_2\right\rangle$ one, and its wave function should be qualitatively similar to the one of the $0^{++}$ glueball, since both states would have $\left\langle \bm L^{\, 2}\right\rangle=0$. With helicity-1 gluons however, $\left\langle \bm L^{\, 2}\right\rangle=4$ in the tensor channel~\cite{gluheli}, and the wave function should thus be qualitatively different from the one of the scalar glueball. The data of Ref.~\cite{gluwf} are unfortunately not accurate enough to make a definitive comparison between the scalar and tensor glueball wave functions, although their behavior seems to be rather different. We hope that new data will be available in the future, in order to further check the relevance of the transverse gluon picture.     
\acknowledgments
The author thanks the F.R.S.-F.N.R.S. for financial support, and Dr Claude Semay for valuable discussions about this work.


\begin{thebibliography}{99}

\bibitem{review} E. Klempt and A. Zaitsev, Phys. Rept. \textbf{454}, 1 (2007), and references therein.
\bibitem{Ochs} W.~Ochs, Nucl.\ Phys.\ Proc.\ Suppl.\  {\bf 174}, 146 (2007).
\bibitem{lattice1} C. J. Morningstar and M. Peardon, Phys. Rev. D
\textbf{60}, 034509 (1999); Y. Chen \textit{et al.},
Phys. Rev. D \textbf{73}, 014516 (2006).
\bibitem{latti4} H. B. Meyer and M. J. Teper, Phys. Lett. B \textbf{605}, 344 (2005). 
\bibitem{szcz} A. Szczepaniak, E. S. Swanson, C.-R. Ji and S. R. Cotanch, Phys. Rev. Lett. \textbf{76}, 2011 (1996)]; A. Szczepaniak and E. S. Swanson, Phys. Lett. B \textbf{577}, 61 (2003).
\bibitem{ads} R.~C.~Brower, S.~D.~Mathur and C.~I.~Tan, Nucl.\ Phys.\  B {\bf 587}, 249 (2000).
\bibitem{glum0} T. Barnes, Z. Phys. C \textbf{10}, 275 (1981); A. B. Kaidalov and Yu. A. Simonov, Phys. Lett. B \textbf{636}, 101 (2006); F. Brau and C. Semay, Phys. Rev. D \textbf{70}, 014017
(2004); F. Buisseret, Phys. Rev. C \textbf{76}, 025206 (2007).
\bibitem{glumm} J.~M.~Cornwall and A.~Soni, Phys. Lett. B {\bf 120}, 431 (1983); W.-S. Hou and A. Soni, Phys. Rev. D \textbf{29}, 101 (1984); W.-S. Hou, C.-S. Luo and G.-G. Wong, Phys. Rev. D \textbf{64}, 014028 (2001). 
\bibitem{vrev}  V.~Mathieu, N.~Kochelev and V.~Vento, to appear in Int. J. Mod. Phys. E [arXiv:0810.4453].
\bibitem{glub} N. Boulanger, F. Buisseret, V. Mathieu and C. Semay, Eur. Phys. J. A \textbf{38}, 317 (2008).
\bibitem{gluheli} V. Mathieu, F. Buisseret and C. Semay, Phys. Rev. D {\bf 77}, 114022 (2008).
\bibitem{lhade}  T.~A.~Lahde, C.~J.~Nyfalt and D.~O.~Riska, Nucl.\ Phys.\  A {\bf 674}, 141 (2000).
\bibitem{lat0} G. S. Bali, Phys. Rep. \textbf{343}, 1 (2001).
\bibitem{koma} Y.~Koma and M.~Koma, Nucl.\ Phys.\  B {\bf 769}, 79 (2007).
\bibitem{glue2} F. Buisseret and C. Semay, Eur. Phys. J. A \textbf{33}, 87 (2007). 
\bibitem{lat} P. de Forcrand and K.-F. Liu, Phys. Rev. Lett.
\textbf{69}, 245 (1992).
\bibitem{gluwf} M. Loan and Y. Ying, Prog. Theor. Phys. \textbf{116},
169 (2006).
\bibitem{0pp} G. B. West, Phys. Rev. Lett. \textbf{77}, 2622 (1996). 
\bibitem{tepernew} H.~B.~Meyer, JHEP {\bf 0901}, 071 (2009).
\bibitem{loannew}  M.~Loan, Eur.\ Phys.\ J.\  C {\bf 54}, 475 (2008). 
\bibitem{jaco} M. Jacob and G. C. Wick, Ann. Phys. \textbf{7}, 404 (1959).
\bibitem{lag1} F. Buisseret and C. Semay, Phys. Rev. E \textbf{75}, 026705 (2007).
\bibitem{baye} D. Baye and P.-H. Heenen, J. Phys. A \textbf{19}, 2041
(1986); C. Semay, D. Baye, M. Hesse and B. Silvestre-Brac,
Phys. Rev. E \textbf{64}, 016703 (2001).
\bibitem{cas} G. S. Bali, Phys. Rev. D \textbf{62}, 114503 (2000); 
S. Kratochvila and P. de Forcrand, Nucl. Phys. B
\textbf{671}, 103 (2003); P. Bicudo, M. Cardoso and O. Oliveira, Phys. Rev. D \textbf{77}, 091504(R) (2008); M. Cardoso and P. Bicudo, Phys. Rev. D \textbf{78}, 074508 (2008).
\end{thebibliography}
\end{document}